\documentclass[conference]{IEEEtran}
\IEEEoverridecommandlockouts

\usepackage{cite}
\usepackage{amsmath,amssymb,amsfonts}
\usepackage{algorithmic}
\usepackage{graphicx}
\usepackage{textcomp}
\usepackage{xcolor}
\usepackage{adjustbox}
\usepackage{enumitem}
\usepackage[linesnumbered,ruled,vlined]{algorithm2e}
\SetKwInput{KwInput}{Input}                
\SetKwInput{KwOutput}{Output}  
\nonstopmode
\begin{document}

\title{Analysis of Label-Flip Poisoning Attack \\ on Machine Learning Based Malware Detector}

\author{\IEEEauthorblockN{ Kshitiz Aryal}
\IEEEauthorblockA{\textit{Department of Computer Science} \\
\textit{Tennessee Technological University}\\
Cookeville, TN, USA \\
karyal42@tntech.edu}
\and
\IEEEauthorblockN{ Maanak Gupta}
\IEEEauthorblockA{\textit{Department of Computer Science} \\
\textit{Tennessee Technological University}\\
Cookeville, TN, USA \\
mgupta@tntech.edu}
\and
\IEEEauthorblockN{Mahmoud Abdelsalam}
\IEEEauthorblockA{\textit{Department of Computer Science} \\
\textit{North Carolina A\&T State University}\\
Greensboro, NC, USA \\
mabdelsalam1@ncat.edu
}
}

\maketitle

\begin{abstract}
With the increase in machine learning (ML) applications in different domains, incentives for deceiving these models have reached more than ever. As data is the core backbone of ML algorithms, attackers shifted their interest towards polluting the training data itself. Data credibility is at even higher risk with the rise of state-of-art research topics like open design principles, federated learning, and crowd-sourcing. Since the machine learning model depends on different stakeholders for obtaining data, there are no existing reliable automated mechanisms to verify the veracity of data from each source.

Malware detection is arduous due to its malicious nature with the addition of metamorphic and polymorphic ability in the evolving samples. ML has proven to solve the zero-day malware detection problem, which is unresolved by traditional signature-based approaches. The poisoning of malware training data can allow the malware files to go undetected by the ML-based malware detectors, helping the attackers to fulfill their malicious goals. A feasibility analysis of the data poisoning threat in the malware detection domain is still lacking. Our work will focus on two major sections: training ML-based malware detectors and poisoning the training data using the label-poisoning approach. We will analyze the robustness of different machine learning models against data poisoning with varying volumes of poisoning data.
\end{abstract}


\section{Introduction}
\label{sec: Intro}
Machine Learning (ML) techniques have been emerging rapidly, providing computational intelligence to various applications. The ability of machine learning to generalize to unseen data has paved its way from labs to the real world. It has already gained unprecedented success in many fields like image processing~\cite{6248110,MULTICOLUMN}, natural language processing~\cite{review,doi:10.1126/science.aaa8685}, recommendation systems used by Google, YouTube and Facebook, cybersecurity~\cite{TSAI200911994,6735264}, robotics~\cite{5152577}, drug research~\cite{manicavasaga2022drug, dhakal2022artificial}, and many other domains. ML-based systems are achieving unparalleled performance through modern deep neural networks bringing revolutions in AI-based services. Recent works have shown significant achievements in fields like self-driving cars and voice-controlled systems used by tech giants like autopilot in Tesla, Apple Siri, Amazon Alexa, and Microsoft Cortana. With machine learning being applied to such critical applications, continuous security threats are never a bombshell. In addition to traditional security threats like malware attack~\cite{6169963}, phishing~\cite{7813778}, man-in-the-middle attack~\cite{4768661}, denial-of-service~\cite{601338}, SQL injection~\cite{Halfond2006ACO}, adversaries are finding novel ways to sneak into ML models~\cite{yilmaz2019expansion}.

Data poisoning and evasion attacks~\cite{8553214, Kreuk2018AdversarialEO, Demetrio2019, Suciu2019Malware, aryal2021survey} are the latest menaces against the security of machine learning models. Poisoning attacks enable attackers to control the model's behavior by manipulating a model's data, algorithms, or hyperparameters during the model training phase. On the other hand, an evasion attack is carried out during the test time by manipulating the test sample. Adversaries can craft legitimate inputs imperceptible to humans but force models to make wrong predictions. Szegedy et al.~\cite{szegedy2014intriguing} discovered the vulnerability of deep learning architecture against adversarial attacks, and ever since, there have been several major successful adversarial attacks against machine learning architectures~\cite{dinakarrao2019adversarial,hu2017generating}.
Sophisticated attackers are motivated by very high incentives to manipulate the result of the machine learning models. With the current data scale with which machine learning models are trained, it is impossible to verify each data point individually.

In most scenarios, it is unlikely that an attacker gets access to training data. However, with many systems adopting online learning~\cite{8187324}, crowd-sourcing~\cite{10.1145/2348543.2348580} for training data, open design principles, and federated learning, poisoning attacks already pose a serious threat to ML models \cite{bonawitz2019federated}. There have been instances \cite{the_guardian_2016} when big companies have been compromised by a data poisoning attack. Malware public databases like VirusTotal\footnote{https://www.virustotal.com/}, which rely on crowdsourced malware files for training its algorithm, can be poisoned by attackers while Google's mail spam filter can be thrown out of track by wrong reporting of spam emails. 

\begin{figure}[!t]
    \centering
    \includegraphics[scale=0.5]{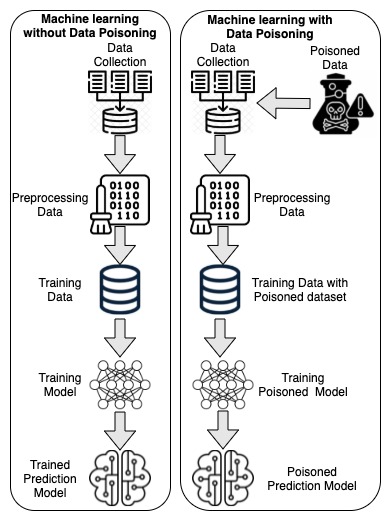}
    \vspace{-4mm}
    \caption{General architecture for Poisoning Machine Learning Models}
    \label{fig:poison}
    \vspace{-6mm}
\end{figure}

Data poisoning relates to adding training data that either leaves a backdoor on the model or negatively impacts the model's performance. Figure \ref{fig:poison} shows the architecture of the poisoning attack. In the given figure, the addition of poisoned data in the training bag forces the model to learn and predict so that attackers benefit from it. This type of  poisoning is not limited to particular domains but has extended across all ML applications. Label flipping attack is carried out to flip the prediction of machine learning detectors. Among all the existing approaches, we chose one of the simplest poisoning techniques called \textit{label poisoning}. We swap the existing training data labels in label poisoning to check the ML models' robustness. 

In this work, we perform a comparative analysis of different machine learning-based malware detectors' robustness against label-flipping data poisoning attacks. Unlike the existing approaches, we are demonstrating the impact of simple label-switching data poisoning in different malware detectors. We will first train eight different ML models widely used to detect malware, namely Stochastic Gradient Descent (SGD), Random Forest (RF), Logistic Regression (LR), K-Nearest Neighbor Classifier (KNN), Linear Support Vector Machine (SVM), Decision Tree (DT), Perceptron, and Multi-Layer Perceptron (MLP). This will be followed by poisoning 10\% and 20\% of training data by flipping the label of data samples. All of the models are retrained after data poisoning, and the performance of each model is evaluated. The major contributions of this paper are as follows.
\begin{itemize}[leftmargin = 3mm]
    \item We taxonomize the existing data poisoning attacks on machine learning models in terms of domains, approaches, and targets.
    
    \item We provide threat modeling for adversarial poisoning attacks against malware detectors. The threat is modeled in terms of the attack surface, the attacker's knowledge, the attacker's capability, and adversarial goals. 
    
    \item We train eight different machine learning-based malware detectors from malware data obtained from VirusTotal and VirusShare\footnote{https://virusshare.com/}. We compare the performance of these malware detectors with training and testing data in terms of accuracy, precision, and recall.
    
    \item Finally, we show a simple label-switching approach to poison the data without any knowledge of training models. The performance of malware detectors is analyzed while poisoning 10\% and 20\% of the total training data. 
\end{itemize}

The rest of the paper is organized as follows. The existing literature for data poisoning attacks in different domains, including malware, is discussed in Section \ref{sec:literature}. Section \ref{sec:threat_model} provides the threat modeling for data poisoning attacks. An overview of ML algorithms that are used to train the malware detector in this paper is discussed in Section \ref{sec:Overview_of_ML}. Section \ref{sec:Methodology} discusses experimental methodology elaborating on the algorithm and the testbed used for the experiment. The evaluation and discussion on the performed experiments are given in Section \ref{sec:Evaluation}. Finally, Section \ref{sec:Conclusion} concludes this work.  

\begin{figure}[!t]
    \centering
    \includegraphics[width = .8\linewidth]{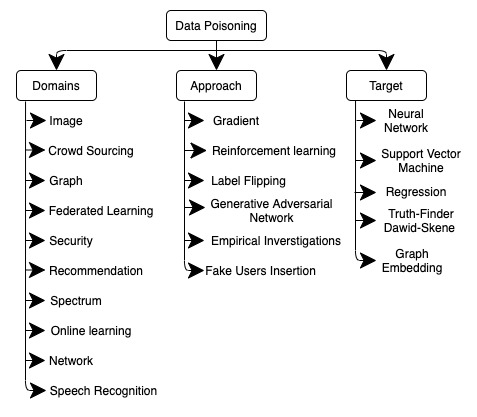}
    \vspace{-4mm}
    \caption{Taxonomy of poisoning attack on attack domain, approach and target}
    \label{fig:taxonomy}
    \vspace{-6mm}
\end{figure}

\begin{table*}[!t]
    \caption{Data Poisoning Attacks}
    \centering
    \def\arraystretch{1.2}
    \begin{adjustbox}{width=0.7\textwidth,center}
    \begin{tabular}{|p{3cm}||c|c|c|c|c|c||c|c|c|c|c||c|c|c|c|c|c|}
    \hline
    &\multicolumn{5}{c}{\textbf{Domains}}& &\multicolumn{4}{c}{\textbf{Approach}}&&\multicolumn{4}{c}{\textbf{Target}}&\\
    \hline
    \textbf{Publications} &
    \rotatebox[origin=c]{90}{\textbf{Image}} & 
    \rotatebox[origin=c]{90}{\textbf{\parbox{2cm}{\centering Crowd\\Sourcing}}} &
    \rotatebox[origin=c]{90}{\textbf{Graph}} &
    \rotatebox[origin=c]{90}{\textbf{\parbox{2.5cm}{\center Federated\\Learning}}} &
    \rotatebox[origin=c]{90}{\textbf{Security}} &
    \rotatebox[origin=c]{90}{\textbf{\parbox{2.5cm}{\centering Online\\Learning}}} &
     \rotatebox[origin=c]{90}{\textbf{Gradient}} &
     \rotatebox[origin=c]{90}{\textbf{\parbox{2.5cm}{\centering Reinforcement\\Learning}}} &
     \rotatebox[origin=c]{90}{\textbf{\parbox{2.5cm}{\centering Label\\Flipping}}} &
     \rotatebox[origin=c]{90}{\textbf{GAN}}&
     \rotatebox[origin=c]{90}{\textbf{Others}} &
     \rotatebox[origin=c]{90}{\textbf{\parbox{2.5cm}{\centering Neural\\Network}}} &
     \rotatebox[origin=c]{90}{\textbf{SVM}} &
     \rotatebox[origin=c]{90}{\textbf{Regression}} &
     \rotatebox[origin=c]{90}{\textbf{\parbox{2.5cm}{\centering Graph\\Embedding}}} &
     \rotatebox[origin=c]{90}{\textbf{Customized}}\\
     \hline
     Shafahi et al.~\cite{shafahi2018poison}&$\surd$&&&&&&&&&&$\surd$&$\surd$&&&&\\
     \hline
     Liu et al.~\cite{liu2019unified}&$\surd$&&$\surd$&&&&$\surd$&&&&&&&&&$\surd$\\
     \hline
     Cao et al.~\cite{Cao_pois_feder}&&&&$\surd$&&&&&$\surd$&&&$\surd$&&&&\\
     \hline
     Shen et al.~\cite{tensorclog_shen}&$\surd$&&&&&&$\surd$&&&&&$\surd$&&&&\\
     \hline
     Zhang et al.~\cite{zhang_federated}&&&&$\surd$&&&&&&$\surd$&&$\surd$&&&&\\
     \hline
     Jiang et al.~\cite{Jiang_poison}&$\surd$&&&&&&$\surd$&&&&&&&$\surd$&&\\
     \hline
     Kwon et al.~\cite{Kwon_sel_poison}&$\surd$&&&&&&&&&&$\surd$&$\surd$&&&&\\
     \hline
     Zhang et al.~\cite{ijcai2019-674}&&&$\surd$&&&&&&&&$\surd$&&&&$\surd$&\\
     \hline
     Bagdasaryal et al.~\cite{bagdasaryan2019backdoor}&$\surd$&&&$\surd$&&&&&&&$\surd$&$\surd$&&&&\\
     \hline
     Li et al.~\cite{Li_partial_poison}&&$\surd$&&&&&&$\surd$&&&&&&&&$\surd$\\
     \hline
     Sasaki et al.~\cite{sasaki2019embedding}&&&&&$\surd$&&$\surd$&&&&&&&$\surd$&&\\
     \hline
     Zhang et al.~\cite{zhang2019online}&$\surd$&&&&&$\surd$&&$\surd$&&&&&&$\surd$&&\\
     \hline
     Lovisotto et al.~\cite{Lovisotto_2020}&$\surd$&&&&&&&&&&$\surd$&$\surd$&&&&\\     
      \hline
     Li et al.~\cite{li2021backdoor} &&&&&$\surd$&&&&$\surd$&&$\surd$&$\surd$&&&&\\
      \hline
     Kravchik et al.~\cite{kravchik2021poisoning}&&&&&$\surd$&&$\surd$&&&&&$\surd$&&&&\\
     \hline
     \textbf{This Work} &&&&&$\surd$&&&&$\surd$&&&$\surd$&$\surd$&$\surd$&&\\
     \hline
  \end{tabular}
  \end{adjustbox}
    \footnotesize\textit{\textbf{Domains}:Poisoning domain for crafted attack, \textbf{Approach}: Approach to poison the training data}, \textbf{Target}: Target of poisoning attack
    \label{tab:survey}
   \vspace{-5mm}
\end{table*}

\section{Literature Review}
\label{sec:literature}

Data poisoning attacks have been used against the machine learning domain for a long time. The existing literature on data poisoning attacks can be taxonomized in terms of attack domains, approach, and the target (victim), as illustrated in Figure \ref{fig:taxonomy}. The recently trending technologies like crowd-sourcing and federated learning are always vulnerable as the veracity of individual data can never be verified. The recent victims of poisoning attacks have spread in security, network, and speech recognition domains. We also classified the major approaches that are taken to produce or optimize the poisoning attacks in Figure \ref{fig:taxonomy}.  The existing data poisoning approaches have targeted almost all the machine learning algorithms ranging from traditional algorithms like regression to modern deep neural network architectures.

Table \ref{tab:survey} summarizes the existing literature on poisoning attacks. Biggio et al.~\cite{biggio2013poisoning} attacked a support vector machine using gradient ascent. To make poisoning attacks closer to the real world, Yang et al.~\cite{yang2017generative} used a generative adversarial network with an autoencoder to poison deep neural nets. Gongalez et al.~\cite{munozgonzalez2017poisoning} extended poisoning from binary learning to multi-class problems. Shafahi et al.~\cite{shafahi2018poison} proposed a targeted clean label poisoning attack on neural networks using an optimization-based crafting method. Shen et al.~\cite{tensorclog_shen} performed an imperceptible poisoning attack on a deep neural network by clogging the back-propagation from gradient tensors during training while also minimizing the gradient norm. Jiang et al.~\cite{Jiang_poison} performed a flexible poisoning attack against linear and logistic regression. Kwon et al.~\cite{Kwon_sel_poison} could selectively poison particular classes against deep neural networks. Cao et al.~\cite{Cao_pois_feder} proposed a distributed label-flipping poisoning approach to poison the DL model in federated architecture. Miao et al.~\cite{disguise_Miao} poisoned Dawid-Skene~\cite{10.2307/2346806} model by exploiting the reliability degree of workers. Fang et al.~\cite{Fang_2018} proposed a poisoning attack against a graph-based recommendation system by maximizing the hit ratio of target items using fake users.

In the given Table \ref{tab:survey}, we can observe that only a handful of works have been carried out in the security domain. Sasaki et al.~\cite{sasaki2019embedding} proposed an attack framework for backdoor embedding, which prevented the detection of specific types of malware. They generated poisoning samples by solving an optimization problem and tested it against a logistic regression-based malware detector. To poison the Android malware detectors, Lie et al.~\cite{li2021backdoor} experimented backdoor poisoning attack against Drebin~\cite{arp2014drebin}, DroidCat~\cite{cai2018droidcat}, MamaDroid~\cite{mariconti2016mamadroid} and DroidAPIMiner~\cite{aafer2013droidapiminer}.  Kravchik et al.~\cite{kravchik2021poisoning} attacked the cyber attack detectors deployed in the industrial control system. The back gradient optimization techniques used to pollute the training data successfully poison the neural network-based model. These works have focused their approach on some algorithm testing against some defense mechanism. However, none of the works compared the feebleness of multiple algorithms against data poisoning attacks. \textit{In this work, we demonstrate the effectiveness of label switch poisoning of the training data against eight machine learning algorithms widely used in malware detectors. }

\section{Threat Model: Know The Adversary}
\label{sec:threat_model}

\begin{figure}[!t]
    \centering
    \includegraphics[width = \linewidth]{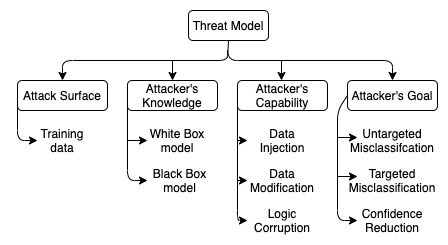}
    \vspace{-4mm}
    \caption{Threat model for poisoning attack}
    \label{fig:threat}
  \vspace{-6mm}
\end{figure}

All security threats are defined in terms of their goals and attack capabilities. Modeling the threat allows for identifying and better understanding the risk arriving with a threat. A poisoning attack is performed by manipulating the training data either at the initial learning or incremental learning period. The threat model of a poisoning attack reflects the attacker's knowledge, goal, capabilities, and attack surface, as shown in Figure~\ref{fig:threat}.

%\subsection{Attack Surface}
\noindent
\textbf{Attack Surface:} 
Attack surface denotes how the adversary attacks the model under analysis. Machine learning algorithms require data to pass through different stages in the pipeline, and each stage offers some kind of vulnerability. In this work, we are only concerned about poisoning attacks which make the training data an attack surface. 

%\subsection{Attacker's Knowledge}
\noindent
\textbf{Attacker's Knowledge:} 
The attacker's knowledge is the amount of information about the model under attack that an attacker has. Based on the amount of knowledge of the attacker, the poisoning approach is determined. Attacker's knowledge can be broadly classified into the two following categories:
\begin{itemize}[leftmargin = 3mm]
    \item \textit{White box model:} In the white box model, an attacker has complete information about the underlying target model, such as the algorithm used, training data, hyper-parameters, and gradient information. It's easier to carry out a white box attack due to the information available that helps the attacker to create a worst-case scenario for the target model. 
    \item \textit{Black box model:} In the black-box model, an attacker only has information about the model's input and output. An attacker has no information about the internal structure of the model. Black-box models can also be divided further into complete black-box models and gray-box models. In the gray box model, the model's performance for each input the attacker provides can be known. As such, the gray box attack is considered to be relatively easier than the complete black box model. 
\end{itemize}
In this paper, we perform a black box attack on different malware detection models. Our experiments will prove the vulnerability of these models to random label poisoning attacks without having any information about the models.

%\subsection{Attacker's Capability}
\noindent
\textbf{Attacker's Capability:}
The attacker's capability represents the ability of an adversary to manipulate the data and model in different stages of the ML pipeline. It defines the sections that can be manipulated, the mechanism used for manipulation, and constraints to the attacker. Poisoning can be carried out in a well-controlled environment if the attacker has complete information about the underlying model and training data. Attacker capabilities can be classified into the following categories:

\begin{itemize}[leftmargin = 3mm]
    \item \textit{Data Injection:} It is the ability to insert new data into the training dataset, leading machine learning models to learn on contaminated data. 
    \item \textit{Data Modification:} It is the ability to access and modify the training data as well as the data labels. Label flipping is a well-known approach carried out in poisoning attack domains. 
    \item \textit{Logic Corruption:} It is the ability to manipulate the logic of ML models. This ability is out of scope for data poisoning and is considered a model poisoning approach.

\end{itemize}

\noindent
\textbf{Adversarial Goals:}
The attacker's objective is to deceive the ML  model by injecting poisoned data. However, poisoning training data might differ depending on the goals of an attacker. Attacker goals can be categorized as:
\begin{itemize}[leftmargin = 3mm]
    \item \textit{Untargeted Misclassification: } An attacker tries to change the model's output to a value different than the original prediction. Untargeted misclassification is a relatively easier goal for attackers.
    \item \textit{Targeted Misclassification: } An attacker's goal is to add a certain backdoor in the models so that particular samples are classified to a chosen class. 
    \item \textit{Confidence Reduction: } An attacker can also poison training data to reduce the confidence of the machine learning model for a particular prediction. In this approach, changing the classification label is unnecessary, but reducing the confidence score is enough to meet the attacker's goal. 
\end{itemize}
Our paper aims to cause the malware detector models to misclassify. However, since we are dealing with binary classification, it can be considered either targeted or untargeted misclassification.

\section{Overview of Machine Learning Algorithms}
\label{sec:Overview_of_ML}
Almost all of the ML architectures have already been victimized by data poisoning attacks. In this section, we will brief some ML architectures in which we performed data poisoning attacks later in this paper. 

\begin{itemize}[leftmargin = 0 mm]
    \item [] \textbf{Stochastic Gradient Descent: } Stochastic gradient descent (SGD) is derived from the gradient descent algorithm, which is a popular ML optimization technique. A gradient gives the slope of the function and measures the degree of change of a variable in response to the changes of another variable. Starting from an initial value, gradient descent runs iteratively to find the optimal values of the parameters, which are the minimal possible value of the given cost function. In Stochastic Gradient Descent, a few samples are randomly selected in place of the whole dataset for each iteration. The term batch determines the number of samples to calculate each iteration's gradient. In normal gradient descent optimization, a batch is taken to be the whole dataset leading to the problem when the dataset gets big. Stochastic gradient descent considers a small batch in each iteration to lower the computing cost of the gradient descent approach while working with a large dataset.
    
    \item [] \textbf{Random Forest: } A random forest is a supervised ML algorithm that is constructed from an ensemble of decision tree algorithms. Its ensemble nature helps to provide a solution to complex problems. The random forest is made up of a large number of decision trees that have been trained via bagging or bootstrap aggregation. The average mean of the output of constituent decision trees is the random forest's ultimate forecast. The precision of the output improves as the number of decision trees used grows. A random forest overcomes the decision tree algorithm's limitations by eliminating over-fitting and enhancing precision.
    
    \item [] \textbf{Logistic Regression: } The probability for classification problems is modeled using logistic regression, which divides them into two possible outcomes. For classification, logistic regression is an extension of the linear regression model. For regression tasks, linear regression works well; however, it fails to replicate for classification. The linear model considers the class a number and finds the optimum hyperplane that minimizes the distances between the points and the hyperplane. As it interpolates between the points, it cannot be interpreted as probabilities. Because there is no relevant threshold for class separation, logistic regression is applied. It is a widely used classification algorithm due to its ease of implementation and strong performance in linearly separable classes. 
    
    \item [] \textbf{K-Nearest Neighbors (KNN) Classifier: } The KNN algorithm relies on the assumption that similar things exist in close proximity. It is a non-parametric and lazy learning algorithm. KNN does not carry any assumption for underlying data distribution. It does not require training data points for model generation, as all the training data are used in a testing phase. This results in faster training and a slower testing process. The costly testing phase will consume more time and memory. In KNN, K is the number of nearest neighbors and is generally considered odd. KNN, however, suffers from the curse of dimensionality. With increased feature dimension, it requires more data and becomes prone to overfitting.
    
    \begin{algorithm}[!t]
\DontPrintSemicolon
\label{Alg: Data_pois}
\caption{Data Poisoning Algorithm}
    \KwInput{Non-poisoned feature set}
    \KwOutput{Poisoned feature set}
    \KwData{Static features obtained from malware and benign training set}
    \For{all the samples}
    {
    Train the machine learning models and measure the performance
    
    \For{10\% each of Malware and Benign data}
        {
        \If{Training label is not flipped}
            {
            label=Get training label of given data
            
            \If{label==0}
                {
                Flip the label to 1
                }
            \ElseIf{label==1}
                {
                Flip the label to 0
                }
            }
        }
    Train all the models and measure the performance
    
    \For{20\% each of Malware and Benign data}
        {
        \If{Training label is not flipped}
            {
            label=Get training label of given data
            
            \If{label==0}
                {
                Flip the label to 1
                }
            \ElseIf{label==1}
                {
                Flip the label to 0
                }
            }
        }
    Train all the models and measure the performance
    
    }
\end{algorithm}

    \item [] \textbf{Support Vector Machine (SVM): } A support vector machine is a popular supervised ML algorithm applied in both classification and regression tasks. SVM aims to find a hyperplane that classifies the data points. In SVM, there are several possible hyperplanes, and we need to determine the optimal hyperplane that maximizes the margin between the two classes. Hyperplanes are the decision boundary for SVM, where data points near to hyperplane are the support vectors. Due to its effectiveness in high dimensional spaces and memory-efficient properties, it is widely adopted in different domains. 
    
\begin{table*}[!t]
\centering
\caption{Malware Detection Training Result}
%\vspace{-4mm}
\label{Tab:Det_result}
\begin{tabular}{|l|rrrrrrrr|}
\hline
\multicolumn{1}{|c|}{{\textbf{Algorithm}}} & \multicolumn{8}{c|}{\textbf{Clean Data}} 
\\ \cline{2-9} 
\multicolumn{1}{|c|}{}                                    & \multicolumn{4}{c|}{\textbf{Training Data}}                                                                             & \multicolumn{4}{c|}{\textbf{Testing Data}}                                                                              \\ \cline{2-9} 
\multicolumn{1}{|c|}{}                                    & \multicolumn{1}{l|}{\textbf{Accuracy}} & \multicolumn{1}{l|}{\textbf{Precision}} & \multicolumn{1}{l|}{\textbf{Recall}} & \multicolumn{1}{l|}{\textbf{F1}} & \multicolumn{1}{l|}{\textbf{Accuracy}} & \multicolumn{1}{l|}{\textbf{Precision}} & \multicolumn{1}{l|}{\textbf{Recall}} & \multicolumn{1}{l|}{\textbf{F1}} \\ \hline
Stochastic Gradient Descent                               & \multicolumn{1}{r|}{93.41}             & \multicolumn{1}{r|}{92.49}              & \multicolumn{1}{r|}{88.29}  
&\multicolumn{1}{r|}{90.34}      & \multicolumn{1}{r|}{72.98}             & \multicolumn{1}{r|}{58.6}               & \multicolumn{1}{r|}{78.77}  & 67.20                           \\ \hline
Decision Tree                                             & \multicolumn{1}{r|}{99.96}             & \multicolumn{1}{r|}{99.98}              & \multicolumn{1}{r|}{99.91}           & \multicolumn{1}{r|}{99.94} & \multicolumn{1}{r|}{59.65}             & \multicolumn{1}{r|}{44.5}               & \multicolumn{1}{r|}{59.85} &   51.05                          \\ \hline
Random Forest                                             & \multicolumn{1}{r|}{99.97}             & \multicolumn{1}{r|}{99.92}              & \multicolumn{1}{r|}{99.97}           & 
\multicolumn{1}{r|}{99.94}  & \multicolumn{1}{r|}{83.65}             & \multicolumn{1}{r|}{98.82}              & \multicolumn{1}{r|}{54.12} &  69.94                               \\ \hline
Logistic Regression                                       & \multicolumn{1}{r|}{93.2}              & \multicolumn{1}{r|}{92.21}              & \multicolumn{1}{r|}{87.94}           & \multicolumn{1}{r|}{90.02} & \multicolumn{1}{r|}{92.33}             & \multicolumn{1}{r|}{92.24}              & \multicolumn{1}{r|}{85.36} &  88.67                           \\ \hline
KNN Classifier                                            & \multicolumn{1}{r|}{98.38}             & \multicolumn{1}{r|}{97.33}              & \multicolumn{1}{r|}{98.05}           & \multicolumn{1}{r|}{97.69} & \multicolumn{1}{r|}{97.42}             & \multicolumn{1}{r|}{96.38}              & \multicolumn{1}{r|}{96.25} &     96.31                            \\ \hline
Support Vector Machine                                    & \multicolumn{1}{r|}{93.15}             & \multicolumn{1}{r|}{92.44}              & \multicolumn{1}{r|}{87.51}           & \multicolumn{1}{r|}{89.91} & \multicolumn{1}{r|}{92.03}             & \multicolumn{1}{r|}{90.89}              & \multicolumn{1}{r|}{85.94} &        88.34                         \\ \hline
Perceptron                                                & \multicolumn{1}{r|}{90.93}             & \multicolumn{1}{r|}{88.6}               & \multicolumn{1}{r|}{84.91}           & \multicolumn{1}{r|}{86.72} & \multicolumn{1}{r|}{75.39}             & \multicolumn{1}{r|}{60.28}              & \multicolumn{1}{r|}{87.86} &       71.50                          \\ \hline
Multi-Layer Perceptron                                    & \multicolumn{1}{r|}{91.28}             & \multicolumn{1}{r|}{91.07}              & \multicolumn{1}{r|}{83.16}           & \multicolumn{1}{r|}{86.94} & \multicolumn{1}{r|}{71.93}             & \multicolumn{1}{r|}{57.45}              & \multicolumn{1}{r|}{77.66}        & 66.04 \\ \hline
\end{tabular}
%\vspace{-3mm}
\end{table*}

\begin{table*}[!t]
\centering
\caption{Malware Detection Performance with 10\% Poisoning Data}
%\vspace{-4mm}
\label{Tab:Det_result_10P}
\resizebox{.8\textwidth}{!}{%
\begin{tabular}{|l|cccccccc|}
\hline
\multicolumn{1}{|c|}{{\textbf{Algorithm}}} & \multicolumn{8}{c|}{\textbf{10\% Poisoned Data}}                                                                                                                                                                             \\ \cline{2-9} 
\multicolumn{1}{|c|}{}                                    & \multicolumn{4}{c|}{\textbf{Training Data}}                                                                             & \multicolumn{4}{c|}{\textbf{Testing Data}}                                                         \\ \cline{2-9} 
\multicolumn{1}{|c|}{}                                    & \multicolumn{1}{c|}{\textbf{Accuracy}} & \multicolumn{1}{c|}{\textbf{Precision}} & \multicolumn{1}{c|}{\textbf{Recall}} & \multicolumn{1}{c|}{\textbf{F1}} &\multicolumn{1}{c|}{\textbf{Accuracy}} & \multicolumn{1}{c|}{\textbf{Precision}} & \textbf{Recall} &\multicolumn{1}{|c|}{\textbf{F1}} \\ \hline
Stochastic Gradient Descent                               & \multicolumn{1}{c|}{85.12}             & \multicolumn{1}{c|}{82.49}              & \multicolumn{1}{c|}{77.14}           & \multicolumn{1}{c|}{79.73} & \multicolumn{1}{c|}{72.39}             & \multicolumn{1}{c|}{64.23}              & \multicolumn{1}{r|}{61.38} &       62.77     \\ \hline
Decision Tree                                             & \multicolumn{1}{c|}{96.77}             & \multicolumn{1}{c|}{99.44}              & \multicolumn{1}{c|}{92.01}           & \multicolumn{1}{c|}{95.58} & \multicolumn{1}{c|}{51.92}             & \multicolumn{1}{c|}{38.33}              & \multicolumn{1}{c|}{43.98} &   40.96        \\ \hline
Random Forest                                             & \multicolumn{1}{c|}{96.77}             & \multicolumn{1}{c|}{98.92}              & \multicolumn{1}{c|}{92.51}           & \multicolumn{1}{c|}{95.61} & \multicolumn{1}{c|}{80.13}             & \multicolumn{1}{c|}{82.68}              & \multicolumn{1}{c|}{60.22} & 69.68           \\ \hline
Logistic Regression                                       & \multicolumn{1}{c|}{84.51}             & \multicolumn{1}{c|}{82.29}              & \multicolumn{1}{c|}{75.39}           & \multicolumn{1}{c|}{78.69} & \multicolumn{1}{c|}{83.26}             & \multicolumn{1}{c|}{81.06}              & \multicolumn{1}{c|}{72.91} & 76.77           \\ \hline
KNN Classifier                                            & \multicolumn{1}{c|}{89.49}             & \multicolumn{1}{c|}{85.47}              & \multicolumn{1}{c|}{87.1}            & \multicolumn{1}{c|}{86.28} & \multicolumn{1}{c|}{86.59}             & \multicolumn{1}{c|}{83.1}               & \multicolumn{1}{c|}{81.15} & 82.11           \\ \hline
Support Vector Machine                                    & \multicolumn{1}{c|}{84.75}             & \multicolumn{1}{c|}{82.84}              & \multicolumn{1}{c|}{75.42}           & \multicolumn{1}{c|}{78.96} & \multicolumn{1}{c|}{66.99}             & \multicolumn{1}{c|}{63.14}              & \multicolumn{1}{c|}{31.16} &   41.73        \\ \hline
Perceptron                                                & \multicolumn{1}{c|}{77.94}             & \multicolumn{1}{c|}{67.78}              & \multicolumn{1}{c|}{79.69}           & \multicolumn{1}{c|}{73.25} & \multicolumn{1}{c|}{40.16}             & \multicolumn{1}{c|}{25.89}              & \multicolumn{1}{c|}{31} & 73.25              \\ \hline
Multi-Layer Perceptron                                    & \multicolumn{1}{c|}{83.85}             & \multicolumn{1}{c|}{82.72}              & \multicolumn{1}{c|}{72.58}           & \multicolumn{1}{c|}{77.32} & \multicolumn{1}{c|}{83.33}             & \multicolumn{1}{c|}{82.81}              & \multicolumn{1}{c|}{70.74} &  76.30       \\ \hline
\end{tabular}}
%\vspace{-4mm}
\end{table*}
    
    \item [] \textbf{Multi-Layer Perceptron: } The term 'Perceptron' is derived from the ability to perceive, see, and recognize images in a human-like manner. A perceptron machine is based on the neuron, a basic unit of computation, with a cell receiving a series of pairs of inputs and weights. Although the perceptron was originally thought to represent any circuit and logic, non-linear data cannot be represented by a perceptron with only one neuron. Multi-Layer Perceptron was developed to overcome this limitation. In multi-layer perceptron, the mapping between input and output is non-linear. It has input and output layers and several hidden layers stacked with numerous neurons. Because the inputs are merged with the initial weights in a weighted sum and applied to the activation function, the multi-layer perceptron falls under the category of feedforward algorithms. Each linear combination is propagated to the following layer, unlike with a perceptron. 
\end{itemize}

\section{Experimental Methodology}
\label{sec:Methodology}

In this paper, we are using the label-flipping approach to poison the training data. With source class $C_S$ and a target class $C_T$ from a set of classes $C$, the dataset $D_I$ is poisoned. The detailed poisoning performed in the paper is shown in Algorithm \ref{Alg: Data_pois}. We perform a label poisoning attack of different volumes to training data without guiding the poisoning mechanism through machine learning architecture or the loss function. It is an efficient way to showcase the ability of random poisoning to hamper the model's performance. We are training all eight malware detector models three times in total. As illustrated in Algorithm \ref{Alg: Data_pois}, we begin the model training with clean data without adding any noise. After recording the model's performance on clean data, we proceed towards the first stage of poisoning our data. We take 10\% of shuffled training data belonging to each malware and benign class, and we change their labels. We retrain all the models and again measure the performance of the models. We repeat the same operation with 20\% of shuffled training data. The percentage of poisoned data is taken randomly for this experimental purpose, as the goal is to show the impact on the models. The algorithm we followed in carrying out this experiment is not a novel approach but a generic approach to poison the data. 

\subsection{Experimental Environment and Dataset}
\label{sec:Environment}
All the experiments are performed in Google-Colab using Google's GPU. All the implementation will be worked around using python libraries and Scikit-Learn. The training dataset~\cite{kaggle_dataset} is obtained from the Kaggle repository, where data are collected from VirusTotal and VirusShare. The dataset comprises windows PE malware and benign files processed through static executable analysis. The dataset comprises 216,352 files (75,503 benign files and 140,849 malware files) with 54 features.
%which are listed in Figure \ref{fig:features}.
% \begin{figure*}[!t]
%     \centering
%     \includegraphics[scale = 0.75]{Image/features.PNG}
%     \caption{Features used for detection}
%     \label{fig:features}
%     \vspace{-3mm}
% \end{figure*}

\section{Evaluation Results and Analysis}
\label{sec:Evaluation}

\subsection{Data Pre-processing and Transformation}
We begin our experiment by loading data from Kaggle dataset~\cite{kaggle_dataset}. To clean the data, we followed two different approaches. First, we ignored rows that are missing more than 50\% of data, whereas we replaced the null values with the arithmetic mean value of the column for rows with less than 50\% missing values. Second, we normalized the data by scaling the values from 0 to 1. Afterward, 85\% of data were used for training purposes while the remaining 15\% were used for testing purposes. We trained selected eight machine learning models with standard hyper-parameters for each model. We didn't tweak many machine learning parameters to fine-tune the detection accuracy, resulting in significant overfitting in a few models.

\subsection{Performance Indicators}
\label{sec:Performance_Indicator}
We evaluated the malware detectors' performance using the following metrics:

\begin{equation*}
    Accuracy = %\frac{Correct\;Prediction}{Total\;Samples} = 
    \frac{TP+TN}{TP+TN+FP+FN}
\end{equation*}
\begin{equation*}
     Precision = \frac{TP}{TP+FP}, \; Recall = \frac{TP}{TP+FN}
\end{equation*}
\begin{equation*}
    F1\textnormal{-}score = \frac{2*(Precision*Recall)}{Precision + Recall}
\end{equation*}
A positive outcome corresponds to a malware sample, while a negative result corresponds to a benign example. TP, TN, FP, and FN are true positives, true negatives, false positives, and false negatives, respectively. Accuracy is the percentage of correct predictions on the given data. Precision measures the ratio between true positives and all the positives. Recall provides the ability of our model to predict true positives correctly. The F1 score is the harmonic mean, the combination of a classifier's precision and recall.

\begin{table*}[!t]
\centering
\caption{Malware Detection Performance with 20\% Poisoning Data}
%\vspace{-4mm}
\label{Tab:Det_result_20P}
\begin{tabular}{|l|cccccccc|}
\hline
\multicolumn{1}{|c|}{{\textbf{Algorithm}}} & \multicolumn{8}{c|}{\textbf{20\% Poisoned data}}                                                                                                   \\ \cline{2-9} 
\multicolumn{1}{|c|}{}                                    & \multicolumn{4}{c|}{\textbf{Training Data}}                                                                             & \multicolumn{4}{c|}{\textbf{Testing Data}}                                                         \\ \cline{2-9} 
\multicolumn{1}{|c|}{}                                    & \multicolumn{1}{c|}{\textbf{Accuracy}} & \multicolumn{1}{c|}{\textbf{Precision}} & \multicolumn{1}{c|}{\textbf{Recall}} & \multicolumn{1}{|c|}{\textbf{F1}} & \multicolumn{1}{c|}{\textbf{Accuracy}} & \multicolumn{1}{c|}{\textbf{Precision}} & \textbf{Recall} & \multicolumn{1}{|c|}{\textbf{F1}} \\ \hline
Stochastic Gradient Descent                               & \multicolumn{1}{c|}{78.56}             & \multicolumn{1}{c|}{75.65}              & \multicolumn{1}{c|}{70.21}           & \multicolumn{1}{c|}{72.83} & \multicolumn{1}{c|}{62.69}             & \multicolumn{1}{c|}{54.86}              & \multicolumn{1}{c|}{50.72} &     52.71       \\ \hline
Decision Tree                                             & \multicolumn{1}{c|}{96.54}             & \multicolumn{1}{c|}{93.54}              & \multicolumn{1}{c|}{98.34}           & \multicolumn{1}{c|}{95.88} & \multicolumn{1}{c|}{40.26}             & \multicolumn{1}{c|}{34.25}              & \multicolumn{1}{c|}{49.67} &        40.54    \\ \hline
Random Forest                                             & \multicolumn{1}{c|}{96.54}             & \multicolumn{1}{c|}{93.04}              & \multicolumn{1}{c|}{98.94}           & \multicolumn{1}{c|}{95.90} & \multicolumn{1}{c|}{72.8}              & \multicolumn{1}{c|}{68.77}              & \multicolumn{1}{c|}{61.66} & 65.02           \\ \hline
Logistic Regression                                       & \multicolumn{1}{c|}{78.38}             & \multicolumn{1}{c|}{74.3}               & \multicolumn{1}{c|}{72.13}           & \multicolumn{1}{c|}{73.20} & \multicolumn{1}{c|}{77.58}             & \multicolumn{1}{c|}{75.1}               & \multicolumn{1}{c|}{76.78} &    75.93        \\ \hline
KNN Classifier                                            & \multicolumn{1}{c|}{87.41}             & \multicolumn{1}{c|}{82.48}              & \multicolumn{1}{c|}{87.94}           & \multicolumn{1}{c|}{85.12} & \multicolumn{1}{c|}{82.15}             & \multicolumn{1}{c|}{76.16}              & \multicolumn{1}{c|}{82.2} &    79.06         \\ \hline
Support Vector Machine                                    & \multicolumn{1}{c|}{78.58}             & \multicolumn{1}{c|}{74.45}              & \multicolumn{1}{c|}{72.6}            & \multicolumn{1}{c|}{73.51} & \multicolumn{1}{c|}{75.39}             & \multicolumn{1}{c|}{74.74}              & \multicolumn{1}{c|}{60.37} &   66.79         \\ \hline
Perceptron                                                & \multicolumn{1}{c|}{75.16}             & \multicolumn{1}{c|}{68.58}              & \multicolumn{1}{c|}{72.57}           & \multicolumn{1}{c|}{72.57} & \multicolumn{1}{c|}{49.37}             & \multicolumn{1}{c|}{38.28}              & \multicolumn{1}{c|}{38.28} &  38.28     \\ \hline
Multi Layer Perceptron                                    & \multicolumn{1}{c|}{77.6}              & \multicolumn{1}{c|}{75.45}              & \multicolumn{1}{c|}{67.1}            & \multicolumn{1}{c|}{71.03} & \multicolumn{1}{c|}{76.85}             & \multicolumn{1}{c|}{74.81}              & \multicolumn{1}{c|}{65.66} &     69.94       \\ \hline
\end{tabular}
\end{table*}

\begin{figure*}[!t]
    \centering
    \includegraphics[width=0.8\textwidth]{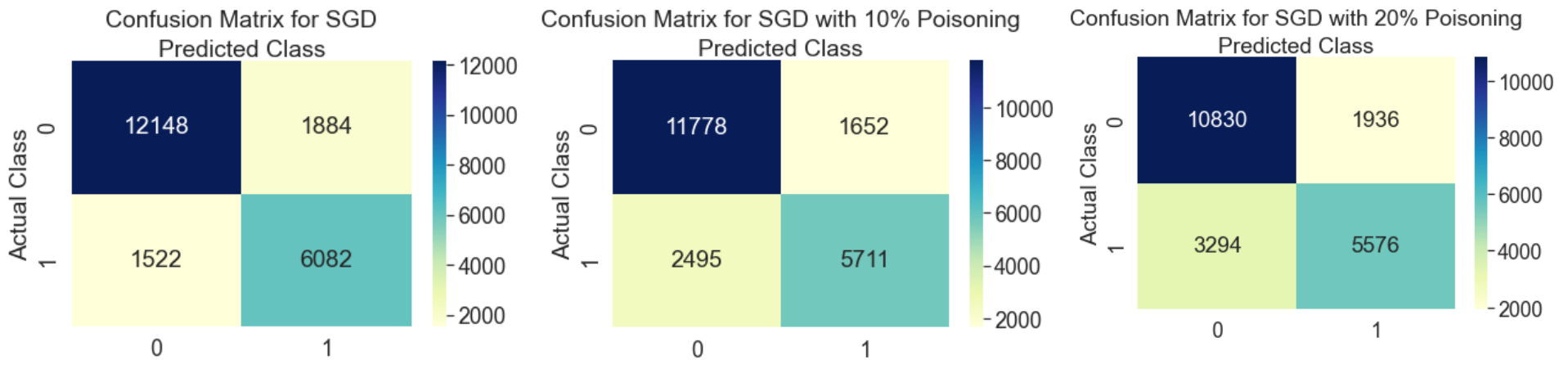}
    \vspace{-5mm}
    \caption{Confusion Matrix for Stochastic Gradient Descent Based Malware Detector }
    \label{fig:SGD}
\end{figure*}

\begin{figure*}[!t]
    \centering
    \includegraphics[width=0.8\textwidth]{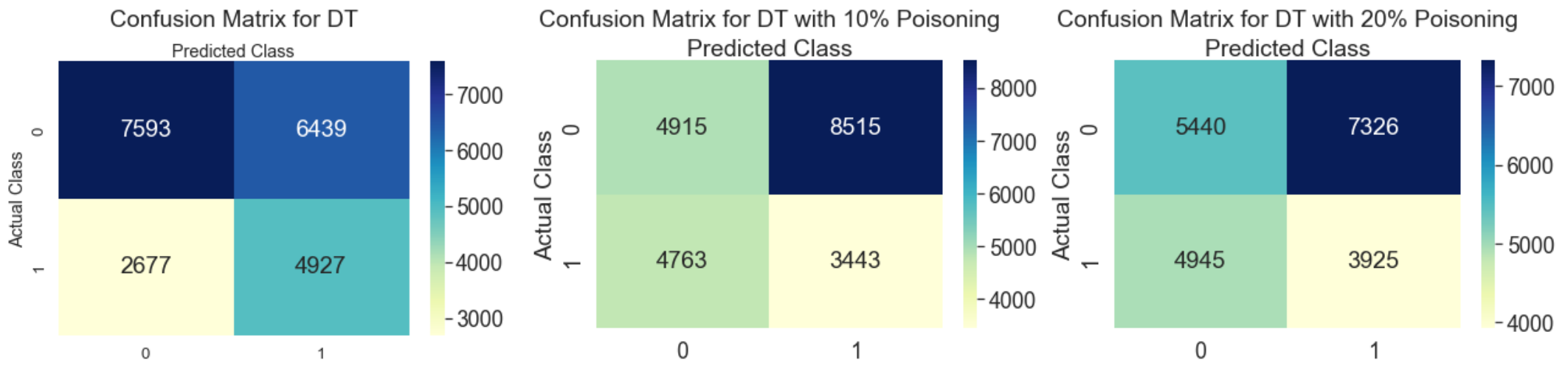}
    \vspace{-5mm}
    \caption{Confusion Matrix for Decision Tree Based Malware Detector}
    \label{fig:DT}
\end{figure*}

\subsection{Results and Discussion}
\label{sec:Results}
Table \ref{Tab:Det_result} shows the accuracy, precision, and recall for training and testing data. Stochastic Gradient Descent, Decision Trees, Random Forest, and Perceptron looked overfitted to training data compared to other models. Since the data volume is a little bit high, decision tree-based classifiers are prone to overfitting problems. We used shallow layer neural networks leading perceptron to overfit in the data. However, classifiers like logistic regression, KNN classifier, and Support Vector Machine have shown the best performance in all three metrics. We have compared the performance of both the training and testing sets as we have only poisoned the training data while preserving the test data from attack. 

\begin{figure*}[!t]
    \centering
    \includegraphics[width=0.8\textwidth]{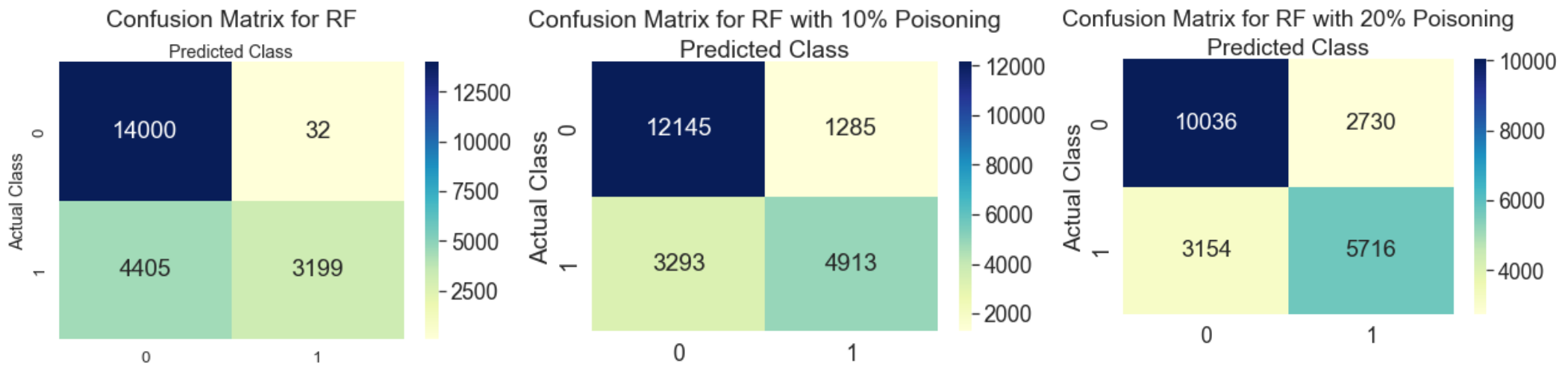}
    \vspace{-5mm}
    \caption{Confusion Matrix for Random Forest-Based Malware Detector}
    \vspace{-2mm}
    \label{fig:RF}
\end{figure*}

\begin{figure*}[!t]
    \centering
    \includegraphics[width=0.8\textwidth]{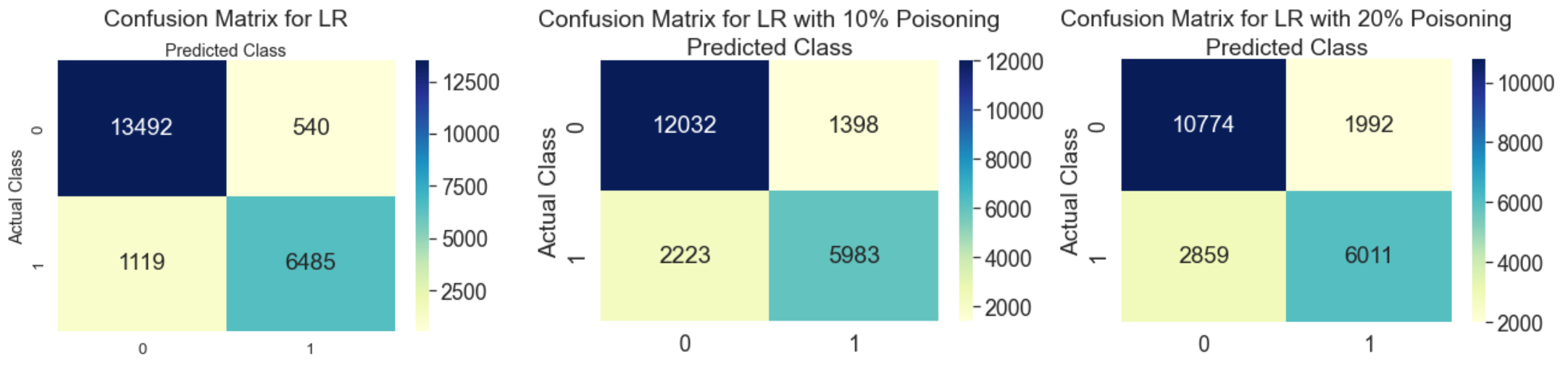}
    \vspace{-5mm}
    \caption{Confusion Matrix for Logistic Regression Based Malware Detector}
   \vspace{-2mm}
    \label{fig:LR}
\end{figure*}

\begin{figure*}[!t]
    \centering
    \includegraphics[width=0.8\textwidth]{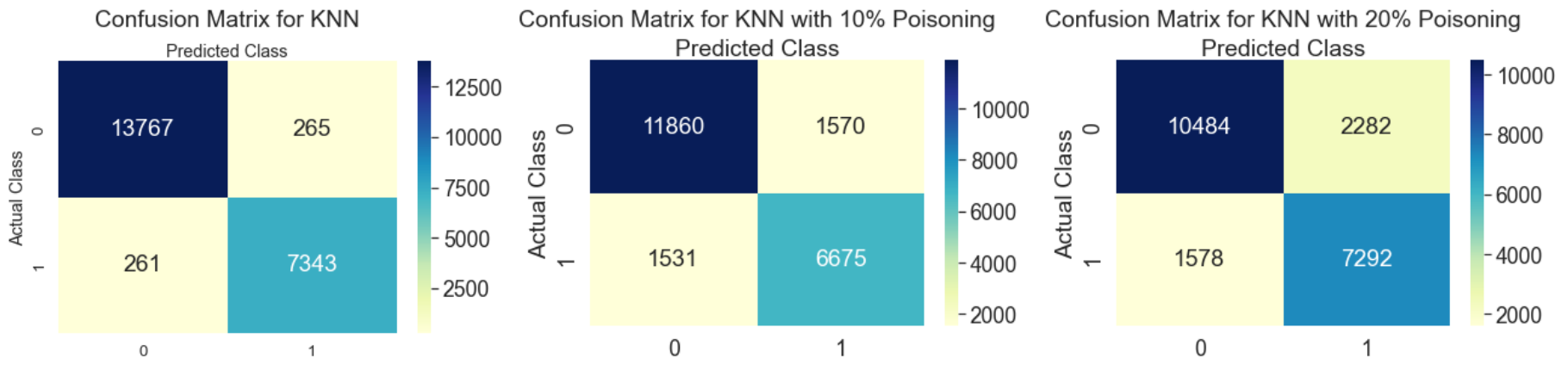}
    \vspace{-5mm}
    \caption{Confusion Matrix for KNN Based Malware Detector}
    \vspace{-2mm}
    \label{fig:KNN}
\end{figure*}

\begin{figure*}[!t]
    \centering
    \includegraphics[width=0.8\textwidth]{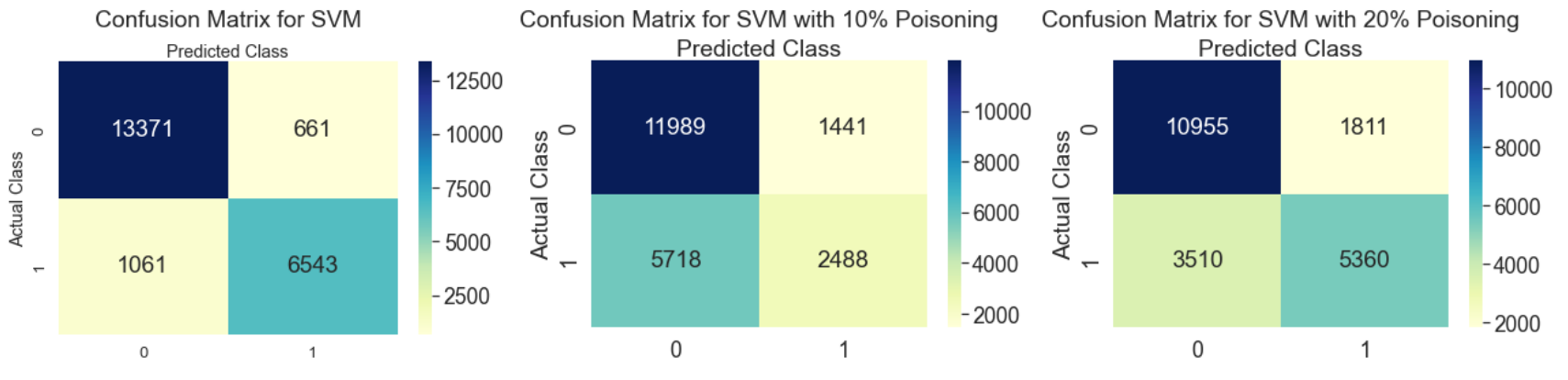}
    \vspace{-5mm}
    \caption{Confusion Matrix for Support Vector Machine-Based Malware Detector}
    \vspace{-2mm}
    \label{fig:SVC}
\end{figure*}

\begin{figure*}[!t]
    \centering
    \includegraphics[width=.8\textwidth]{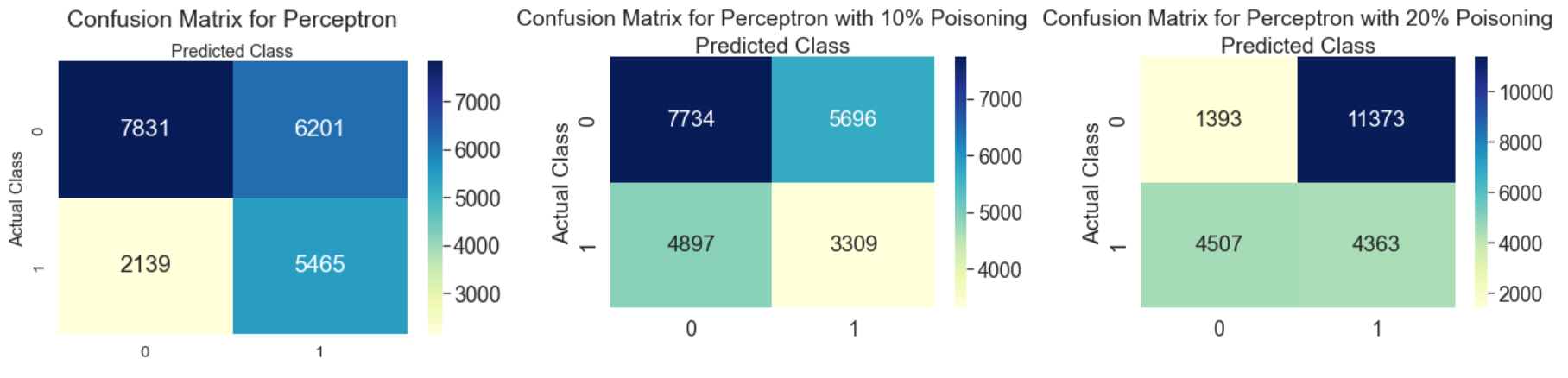}
    \vspace{-5mm}
    \caption{Confusion Matrix for Perceptron Based Malware Detector}
    \label{fig:Percep}
    \vspace{-4mm}
\end{figure*}

\begin{figure*}[!t]
    \centering
    \includegraphics[width=.8\textwidth]{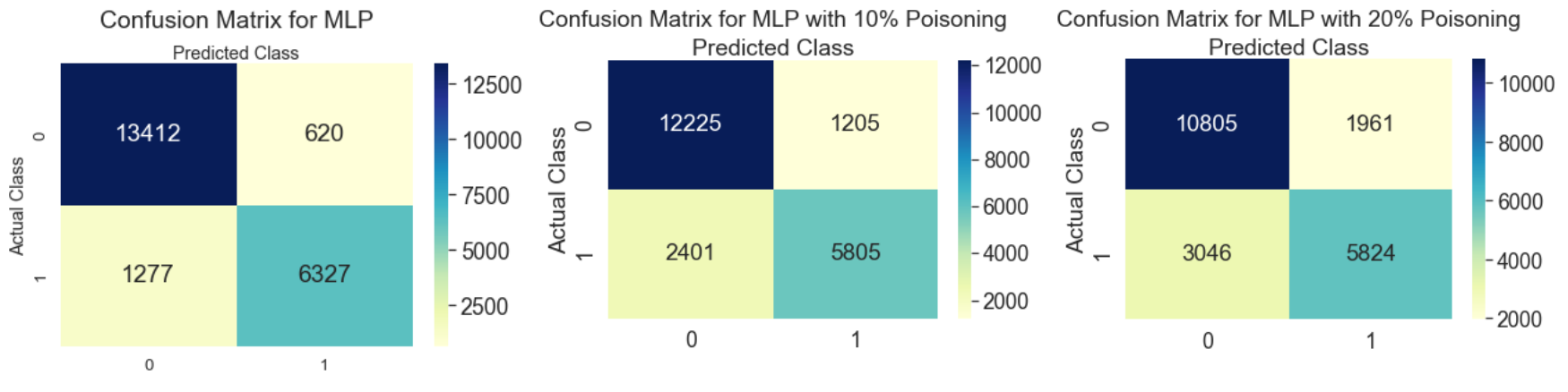}
    \vspace{-5mm}
    \caption{Confusion Matrix for Multi-Layer Perceptron Based Malware Detector}
    \vspace{-2mm}
    \label{fig:MLP}
    \vspace{-4mm}
\end{figure*}

We flipped the labels of 10\% training data as a poisoning attack. On poisoning 10\% of total data, the performance metric for each detector is displayed in Table \ref{Tab:Det_result_10P}. The results show the robustness of decision trees and random forest-based malware detectors compared to other malware detectors. We further poisoned 20\% of total training data to see the impact of increased poisoned data in each model, whose results are shown in Table \ref{Tab:Det_result_20P}. The left-most confusion matrix in each of the figures from Figure \ref{fig:SGD} to Figure \ref{fig:MLP} shows the number of TP, TN, FP, and FN for each classifier on clean data, whereas the middle and right one shows results with 10\% and 20\% poisoning, respectively. In the confusion matrix, label '0' is for malware, and label '1' is for benign samples. The top-left corner in the confusion matrix gives True Positive, the top-right corner gives False Positive, the bottom-left gives False Negative, and the bottom-right corner gives True Negative samples.

\begin{figure*}[!t]
    \centering
    \includegraphics[width=.8\textwidth]{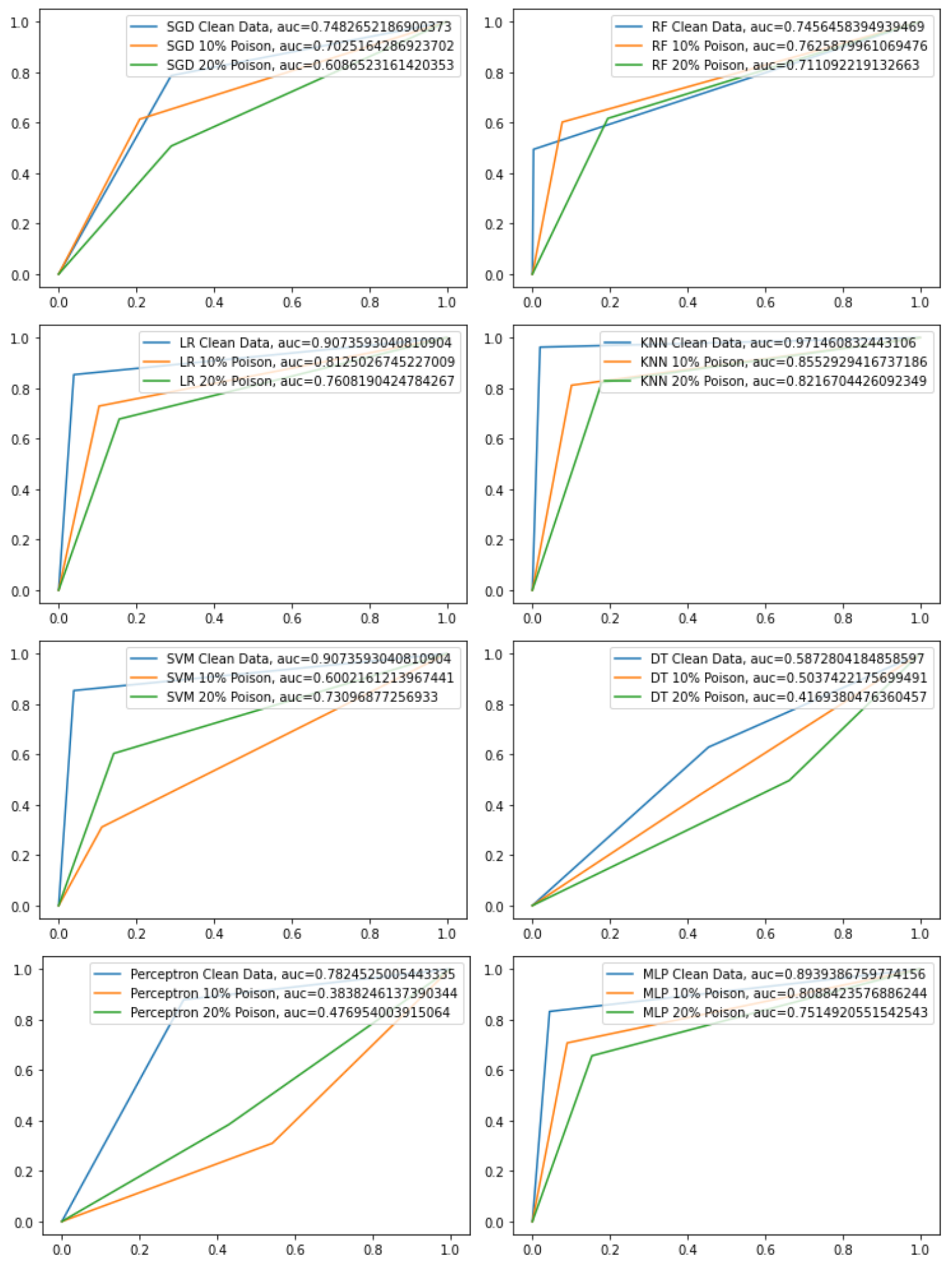}
    \vspace{-5mm}
    \caption{ROC Curve for Malware detectors under Poisoning Environments}
    \label{fig:Ruc}
    \vspace{-6mm}
\end{figure*}

\subsection{Analysis and Observations}
The goal of this work is to show the vulnerability of popular machine-learning models that are used for malware detection. Results in Tables \ref{Tab:Det_result}, \ref{Tab:Det_result_10P} and \ref{Tab:Det_result_20P} reflect the limitations of all the experimented machine learning models even with the basic label poisoning attack. Figure \ref{fig:Ruc} shows the ROC curve, comparing the models' performance on the clean data, 10\% and 20\% poisoned data. In the ROC curve, the blue curve corresponds to the performance of clean data, the orange curve corresponds to 10\% poisoned data, and the green curve corresponds to the 20\% poisoned data. The curve closest to the top-left corner is the one performing best. We can infer from the graph that logistic regression, K-Nearest Neighbors, Support Vector Machine, and Multi-Layer Perceptron are the best models on the clean data. However, the distance between the three curves represents the robustness of the model toward the poisoning attack. If the separation between the curves of clean data and poisoning data is low, it infers that the poisoning attack has a minimal impact on the model's performance. In the ROC graph, we can observe that Random Forest, Logistic Regression, K-Nearest Neighbors, and Multi-Layer Perceptron have their graphs close to each other, proving their robustness against poisoned data. Random Forest's robustness can be attributed to its ensemble nature which helps it to capture better insights about the data. The robustness of logistic regression and K-Nearest Neighbors can be due to the low dimensionality of our training data. 
Further, we can observe the performance of models, like SVM and perceptron, doing better with the 20\% poisoned data than with 10\% poisoned data. The gain in the performance of these models is due to unrestricted data poisoning. Since we are not guiding our poisoning approach according to the models, further adding poisoning data after some threshold point slightly improves the models' performance. 
In the end, even the least sophisticated attacks, like label poisoning, are causing the performance decay of the models to a large extent. This further alerts us toward the catastrophic consequences of more sophisticated attacks like gradients and reinforcement learning. 

\section{Conclusion}
\label{sec:Conclusion}
In this work, we perform a feasibility analysis of label-flip poisoning attacks on ML-based malware detectors. We evaluated eight different ML models that are widely used in malware detection. Spotting the lack of poisoning attacks work in the malware domain, this paper analyses the robustness of ML-based malware detectors against different volumes of poisoned data. We observed the decay in performance of all the models while poisoning 10\% and 20\% of total training data. The significant decrease in the performance of the models shows the severe vulnerability of malware detectors to guided poisoning approaches. We also observed differences in the effect of poisoning attacks across the different models. Our work is carried out within the limited scope of one generic poisoning algorithm and a single malware dataset. There are few future research directions that are clearly visible. The malware detectors can be tested against many advanced poisoning approaches using numerous datasets from the industry. The poisoning can be tested in a more real environment by poisoning the executable files. The research community still lacks exhaustive studies on the vulnerabilities of malware detectors and how to make detectors more robust against these poisoning attacks. 

\bibliographystyle{IEEEtran}
\bibliography{main.bib}

\end{document}